\documentclass[conference]{IEEEtran}
\ifCLASSINFOpdf
\else
   \usepackage[dvips]{graphicx}
\fi
\begin{document}
\title{Design and Analysis on a Cryogenic Current Amplifier with a Superconducting Microwave Resonator}

\author{\IEEEauthorblockN{Yuma Okazaki and Nobu-hisa Kaneko}
\IEEEauthorblockA{National Institute of Advanced Industrial Science and Technology (AIST),\\
Tsukuba, Ibaraki 305-8563, Japan\\
Email: yuma.okazaki@aist.go.jp}
}


%Telephone: (800) 555--1212, Fax: (888) 555--1212}


\maketitle

\begin{abstract}
%\boldmath
We propose a new type of cryogenic current amplifiers, in which low-frequency power spectrum of current can be measured through a measurement of microwave response of a superconducting resonant circuit shunted by a series array of Josephson junctions. From numerical analysis on the equivalent circuit, the numerical value of the input-referred current noise of the proposed amplifier is found to be two orders of magnitude lower than the noise floor measured with the conventional cryogenic current amplifiers based on high-electron-mobility transistors or superconducting quantum interference devices. Our proposal can open new avenues for investigating low-temperature solid-state devices that require lower noise and wider bandwidth power spectrum measurements of current.
\end{abstract}

%\begin{IEEEkeywords}
%cryogenic current amplifier, Josephson junctions, superconducting microwave resonator, current power spectrum.
%\end{IEEEkeywords}

% \fi
\IEEEpeerreviewmaketitle

\section{Introduction}
Precise measurements of power spectrum of current are powerful means for investigating electron transport in solid-state devices with a dilution refrigerator \cite{ade-PicciottoNature97}-\cite{aOkazakiAPL13}. In conventional methods, cryogenic current amplifiers based on  high-electron-mobility transistors (HEMTs) \cite{aDicarloRevSciInstl06}-\cite{aHasisakaRSI14} or superconducting quantum interference devices (SQUIDs) \cite{aJehlRSI1999} are used as the first-stage current amplifier. In these systems, the amplifier's noise severely limits further improvement in the accuracy of measurements; for instance in \cite{aOkazakiAPL13}, it had been reported that the noise of a HEMT-based cryogenic amplifier dominates 94 \% of the total noise floor. To overcome such a limitation, a superconducting resonator connected to Josephson junctions is a promising system, because of its ability to amplify signals with an ultimately lower noise. In previous studies \cite{aVijayRSI09}-\cite{aYamamotoAPL08}, however, the main research interest in such a structure had been amplifying signals in the microwave frequency range, and not amplifying low-frequency current. In contrast, here we propose, based on a similar concept, a design of a cryogenic ``current'' amplifier in low-frequency range and analyze its electrical properties. As a result, we found that the estimated input-referred current noise of the proposed amplifier is two orders of magnitude smaller than that of the conventional cryogenic current amplifier based on HEMTs or SQUIDs.
% because of their ability to amplifying the signal with less noise

Our proposed amplifier consists of a half-wavelength coplanar waveguide resonator as schematically shown in Fig.~1(a). One end of the resonator is coupled to a transmission line used as an input-output port by the capacitor $C_c$, while the other end is shunted by a series array of $N_J$ Josephson junctions (JJ array). To utilize this structure as a current amplifier, a current input port is introduced at the middle of the resonator (`O' in the figure). This input port has two branches; the one is connected to a tunable current source so as to provide a constant bias current $I_B$ to the JJ array; while the other is connected to a device under test (DUT) so that the current $I_S$ from the DUT is fed into the JJ array. In order to avoid an unwanted effect that microwave in the resonator influences the DUT, the input port is placed at the position where the electric field of the relevant resonant mode is zero as depicted by the orange line in the figure. In addition, the input port is inductively isolated from the resonator by the inductors $L_S$ and $L_B$ so as to confine microwave exclusively inside the resonator. Since the kinetic inductance of the Josephson junctions depends on the total current $I = I_B + I_S$ flowing through the junctions, the resonator's mode also depends on the current. Hence measuring frequency response of the resonator allows the current $I_S$ to be measured.

\begin{figure}[h]
 \begin{center}
\includegraphics[]{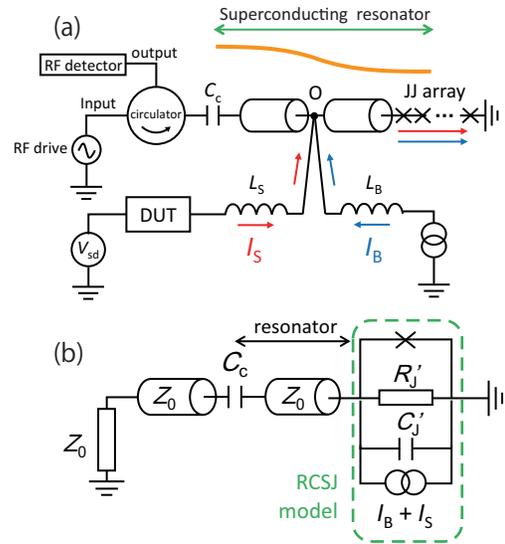}%bb=0 0 253 238
 \end{center}
 \caption[]{(a)Schematic diagram of the proposed cryogenic current amplifier. (b) The equivalent circuit used in the numerical analysis.}
 \label{fig01}
\end{figure}

\section{Numerical model and results}
%Explanation of the simulation model and the parameters toghther with a clear explanation of their validity 
To analyze frequency response of the proposed amplifier, we consider the equivalent circuit shown in Fig.~1(b). In this model, the input-output port and the resonator are represented by transmission lines of a characteristic impedance $Z_0$. Meanwhile, the JJ array is described by a single resistively capacitively shunted junction model (RCSJ model) \cite{aLevinsonPRB84}, since the size of the JJ array, estimated to be tens of micrometers, is negligibly smaller than the typical wave-length of the resonant mode, estimated to be ten millimeters. The net junction resistance $R_J'$ (junction capacitance $C_J'$) of the JJ array is given by $R_J' = N_JR_J$ ($C_J'= C_J/N_J$) with $R_J$ ($C_J$) of the individual junction. Similarly, the net kinetic inductance $L_J'(I)$ of the JJ array, which depends on $I$, is given by $L_J'(I)=N_JL_J(I)$ with $L_J(I) = \hbar / (2e\sqrt{I_C^2 - I^2})$, where $e$, $\hbar$ and $I_C$ are the elementary charge, Planck constant $h$ divided by $2\pi$, and the critical current, respectively \cite{aCastellanos-BeltranAPL07}. In addition, the current $I_B + I_S$ is represented by a current source parallel to the JJ array. We simulate the frequency response of the resonator by calculating the telegrapher's equation with the boundary condition described by the RCSJ model \cite{aJohanssonPRL09}. Throughout the calculation, we use the parameters: $Z_0=50$ $\Omega$, $R_J = 2.4$ k$\Omega$, $C_J = 6$ fF, $I_C = 50$ nA, and $C_C = 10$ fF, determined with reference to the practical experiments \cite{Kakuyanagi}. The resonant frequency of the fundamental mode is designed at 8.1 GHz, because it is easily measured in experiments. On the other hand, $N_J$, $I_B$ and the input microwave power $P_\mathrm{RF}$ are varied so as to find the optimal conditions.

Figures 2(a) and (b) show the frequency response [(a) $S_{11}$ and (b) phase shift $\Delta\phi$] around the resonant frequency $f_0=8.1$ GHz of the fundamental mode with two different current conditions $I=0$ and $I=1$ nA at $N_J = 1$. As expected, the resonant frequency is shifted by a change in $I$. In particular, the phase of the reflected microwave is very sensitive to a change in $I$ as being suggested by the comparison between $d\phi/dI = 0.11$ rad/nA and $dS_{11}/dI \approx 0$ at the on-resonant frequency (the vertical dotted lines in the figures). So, in the present paper, we focus on the situation that the current input to the amplifier is measured through a phase measurement of the reflected microwave. To obtain larger current-to-phase transduction, the bias current $I_B$ should be carefully optimized. Figure 2(c) shows $I_B/I_C$-dependence of current-phase transfer function $\mathrm{d}\phi / \mathrm{d}I$, showing that $\mathrm{d}\phi / \mathrm{d}I$ is enhanced as $I_B$ approaches $I_C$. In what follows, we fix $I_B = 0.95I_C = 45$ nA. Figure 2(d) shows the amount of phase shift $\Delta\phi$ as a function of $I$ with two cases $N_J = 1$ and $N_J = 50$, showing that $I$-dependence of $\Delta\phi$ with $N_J = 50$ is larger than the case with $N_J = 1$ as expected from the fact that the net kinetic inductance $L_J'$ can be enhanced with increasing $N_J$. The linearity in the current-phase relation is also satisfied up to $I\approx 1$ nA in the both cases.

%The frequency response of the resonator
\begin{figure}[!t]
 \begin{center}
\includegraphics[]{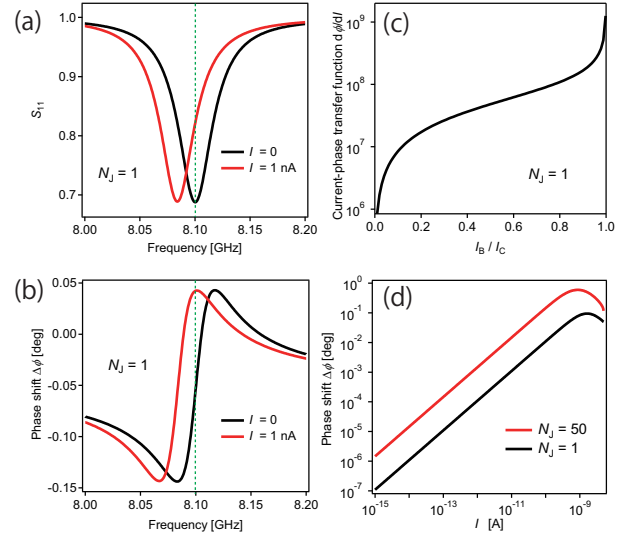}%bb=0 0 253 225
 \end{center}
 \caption[]{(a) and (b) Frequency response of the resonator with $I=0$ and $I=1$ nA. (a) $S_{11}$ and (b) phase shift $\Delta\phi$. (c) The current-phase transfer function $\mathrm{d}\phi / \mathrm{d}I$ as a function of $I_B/I_C$ with $N_J = 1$. (d) $I$-dependence of the phase shift $\Delta\phi$ in the frequency response of the resonator at the on-resonant frequency with $N_J=1$ and $N_J=50$.}
 \label{fig02}
\end{figure}

%Evaluation of the input-referred current noise and its dependence on the input RF power.
The input-referred current noise of the amplifier is estimated by considering two different origins of noise, i.e.~electric noise generated at the JJ array and phase noise added to the frequency response. First, with regard to the former origin, we consider thermal noise and electron shot noise generated at the JJ array. The thermal noise is estimated from the Johnson-Nyquist relation $S_\mathrm{th} = 4k_\mathrm{B}TR_J'$ with $k_\mathrm{B}$ being the Boltzmann constant, while the electron shot noise is estimated from $S_\mathrm{ES} = 2e\langle I_\mathrm{RF}\rangle$ with $\langle I_\mathrm{RF}\rangle$ being the time-averaged current driven by the resonant microwave \cite{aStephenPR69B}. Note here that $I_\mathrm{RF}$, which is a high-frequency excess current carried by quasiparticles, differs from $I_B$ or $I_S$, which is a low-frequency supercurrent carried by the Cooper pairs and does not generate the electron shot noise \cite{aLevinsonPRB84, aStephenPR69A}. Second, with regard to the latter origin, we consider optical shot noise and quantum noise \cite{aJohanssonPRL09}. The optical shot noise, which is associated with stochastic fluctuations in the photon number, is estimated from $(\Delta\phi)^2 = Z_0hf/2V_\mathrm{out}^2$, where $V_\mathrm{out}$ is the voltage amplitude of the reflected microwave. The quantum noise is also estimated from the similar form $(\Delta\phi)^2 = Z_0hf/4V_\mathrm{out}^2$. To quantify the input-referred value, the estimated phase noise should be converted to the corresponding current at the input via $\mathrm{d}\phi / \mathrm{d}I$. We note that the validity of the above estimations was confirmed by checking the agreement of their values with the actual experiments for instance in \cite{aRegalNatPhys08}. Figures 3(b) and (c) show the estimated input-referred current noise $S_\mathrm{I}$ as a function of the input microwave power $P_\mathrm{RF}$ with (b) $N_J = 1$ and (c) $N_J=50$. In the figures, the electron shot noise, optical shot noise and quantum noise are explicitly shown as dashed lines, while the thermal noise is not since its value is negligibly small when the system temperature is $T<100$ mK (typical temperatures in a dilution refrigerator). As expected, the total noise with $N_J = 50$ is lower than that with $N_J = 1$.

\begin{figure}[!t]
 \begin{center}
\includegraphics[]{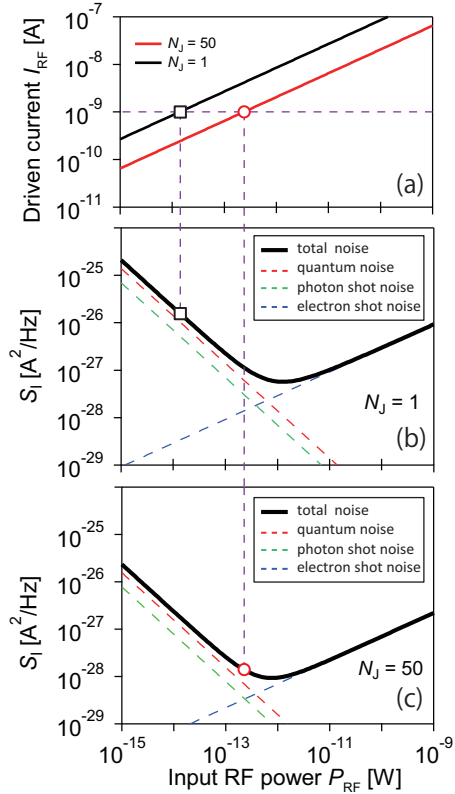}%bb=0 0 253 322
 \end{center}
 \caption[]{(a) An excess current $I_\mathrm{RF}$ driven by the input microwave as a function of its power. (b) and (c) The input refereed current noise of the amplifier with (b) $N_J=1$ and (c) $N_J=50$.}
 \label{fig03}
\end{figure}

Figure 3(b) and (c) show the minimum $S_I$ at $P_\mathrm{RF}\approx 10^{-12}$ W, which reflects the fact that the total noise is comprised of both monotonically decreasing (optical shot noise and quantum noise) and increasing (electron shot noise) components with the input RF power. In order to determine the optimal value of the input RF power, we should take into account the possibility that a larger $I_\mathrm{RF}$ drives the JJ array into the so-called ``voltage state'' and then the amplifier loses its function. To avoid such a situation, the input RF power should be low enough so that the total current $I_S + I_B + I_\mathrm{RF}$ is kept below $I_C$. Since $I_B = 45$ nA in the present analysis has a 5 nA margin to $I_C$ ($=50$ nA), we conclude that $I_\mathrm{RF} < 1$ nA should at least be satisfied. The square (circle) symbol on the trace in Fig.~3(b) [(c)] indicates the RF power at which $I_\mathrm{RF} = 1$ nA in each $N_J$ as indicated by the corresponding symbols in Fig.~3(a). So the RF power should be lower than these points. In the case with $N_J = 50$ [Fig.~3(c)], the input-referred current noise is of the order of $10^{-28}$ - $10^{-27}$ $\mathrm{A^2/Hz}$ at a power range $P_\mathrm{RF}< 10^{-13}$ W. These noises are two orders of magnitude smaller than that with the conventional  cryogenic current amplifiers based on the HEMTs or SQUIDs, whose noise floors are typically of the order of $10^{-26}$ - $10^{-25}$ $\mathrm{A^2/Hz}$ \cite{aOkazakiAPL13, aHasisakaRSI14, aJehlRSI1999}. This improvement in the noise floor allows us to measure the power spectrum of the current from a DUT with an unprecedentedly better signal-to-noise ratio.
%Conclusions

\section{Conclusion}
In conclusion, we designed a new type of cryogenic current amplifiers, in which the power spectrum of current can be measured through a phase measurement of the superconducting resonator shunted by the JJ array. From the numerical analysis on the equivalent circuit, the input-referred current noise of the amplifier is found to be two orders of magnitude smaller than that of the conventional cryogenic amplifiers.

\section*{Acknowledgment}
The authors would like to thank K.~Kakuyanagi, M.~Maezawa, S.~Nakamura, T.~Satoh, Y.~Tokura, H.~Yamaguchi, and H.~Yamamori for fruitful discussions.



%\bibliographystyle{IEEEtran}

\begin{thebibliography}{99}
%\bibitem{rBlanterPR00} Ya.~M.~Blanter, M.~B\"uttiker, ``Shot noise in mesoscopic conductors,'' Phys.~Rep., vol.~336, no.~1-2, pp.~1-166, Sep.~2000.
\bibitem{ade-PicciottoNature97}R.~de-Picciotto , M.~Reznikov, M.~Heiblum, V.~Umansky, G.~Bunin, and D.~Mahalu, ``Direct observation of a fractional charge,'' Nature, vol.~389, no.~11, pp.~162-164, Sep.~1997. %, M.~Reznikov, M.~Heiblum, V.~Umansky, G.~Bunin, and D.~Mahalu
\bibitem{aJehlNature2000}X.~Jehl, M.~Sanquer, R.~Calemczuk, and D.~Mailly, ``Detection of doubled shot noise in short normal-metal/superconductor junctions,'' Nature, vol.~406, no.~6782, pp.~50-53, May 2000.
\bibitem{aOkazakiPRBR13}Y.~Okazaki, S.~Sasaki, and K.~Muraki, ``Shot noise spectroscopy on a semiconductor quantum dot in the elastic and inelastic cotunneling regimes,'' Phys.~Rev.~B, vol.~87, no.~4, p.~041302(R), Jan.~2013.
\bibitem{aOkazakiAPL13}Y.~Okazaki, I.~Mahboob, K.~Onomitsu, S.~Sasaki, and H.~Yamaguchi, ``Quantum point contact displacement transducer for a mechanical resonator at sub-Kelvin temperatures,'' Appl.~Phys.~Lett., vol.~103, no.~19, p.~192105, Nov.~2013.

\bibitem{aDicarloRevSciInstl06}L.~DiCarlo, Y.~Zhang, D.~T.~McClure,C.~M.~Marcus, L.~N.~Pfeiffer, and K.~W.~West, ``System for measuring auto- and cross correlation of current noise at low temperatures,'' Rev.~Sci.~Instrum., vol.~77, no.~7, p.~073906, Jul.~2006.
\bibitem{aHashisakaJPhysConfSer08}M.~Hashisaka, Y.~Yamauchi, S.~Nakamura, S.~Kasai, K.~Kobayashi and T.~Ono, ``Measurement for quantum shot noise in a quantum point contact at low temperatures,'' J.~Phys.:~Conf.~Ser., vol.~109, p.~012013, 2008.
\bibitem{aArakawaAPL13}T.~Arakawa, Y.~Nishihara, M.~Maeda, S.~Norimoto, and K.~Kobayashi, ``Cryogenic amplifier for shot noise measurement at 20 mK,'' Appl.~Phys.~Lett., vol.~103, no.~17, p.~172104, Oct.~2013. 
\bibitem{aHasisakaRSI14}M.~Hashisaka, T.~Ota, M.~Yamagishi, K.~Muraki, and T.~Fujisawa, ``Cross-correlation measurement of quantum shot noise using homemade transimpedance amplifiers,'' Rev.~Sci.~Instrum., vol.~85, no.~5, p.~054704, May.~2014.
\bibitem{aJehlRSI1999}X.~Jehl, P.~Payet-Burin, C.~Baraduc, R.~Calemczuk, and M.~Sanquer, ``Superconducting quantum interference device based resistance bridge for shot noise measurement on low impedance samples,'' Rev.~Sci.~Instrum., Vol.~70, no.~6, Feb.~1999.
%related study aiming at an application to microwave amplifier,
%\bibitem{aEichlerEPJ14}C.~Eichler and A.~Wallraff, ``Controlling the dynamic range of a Josephson parametric amplifier,'' EPJ Quantum Technology 1:2, 2014.
\bibitem{aVijayRSI09}R.~Vijay, M.~H.~Devoret, and I.~Siddiqi, ``The Josephson bifurcation amplifier,'' Rev.~Sci.~Instrum., vol.~80, no.~20, p.~111101, Nov.~2009.
%\bibitem{aHoverAPL12}D.~Hover, Y.-F.~Chen, G.~J.~Ribeill, S.~Zhu, S.~Sendelbach, and R.~McDermott, ``Superconducting low-inductance undulatory galvanometer microwave amplifier,'' Appl.~Phys.~Lett., vol.~ 100, no.~6, p.~063503, Feb.~2012.
%\bibitem{aRibeillJAP11}G.~J.~Ribeill, D.~Hover, Y.-F.~Chen, S.~Zhu, and R.~McDermott, ``Superconducting low-inductance undulatory galvanometer microwave amplifier: Theory,'' J.~Appl.~Phys., vol.~10, no.~10, p.~103901, Nov.~2011.
\bibitem{aCastellanos-BeltranAPL07}M.~A.~Castellanos-Beltran, and K.~W.~Lehnert, ``Widely tunable parametric amplifier based on a superconducting quantum interference device array resonator,'' Appl.~Phys.~Lett., vol.~91, no.~8, p.~083509, Aug.~2007.
\bibitem{aYamamotoAPL08}T.~Yamamoto, K.~Inomata, M.~Watanabe, K.~Matsuba, T.~Miyazaki, W.~D.~Oliver, Y.~Nakamura, and J.~S.~Tsai, ``Flux-driven Josephson parametric amplifier,'' Appl.~Phys.~Lett., vol.~93, no.~4, p.~042510, Jul.~2008.

\bibitem{aJohanssonPRL09}J.~R.~Johansson, G.~Johansson, C.~M.~Wilson, and F.~Nori, ``Dynamical Casimir Effect in a Superconducting Coplanar Waveguide,'' Phys.~Rev.~Lett., vol.~103, no.~14, Oct.~2009.
\bibitem{Kakuyanagi} Kosuke Kakuyanagi (private communication).

\bibitem{aLevinsonPRB84}Y.~Levinson, ``Quantum noise in a current-biased Josephson junction,'' Phys.~Rev.~B, vol.~67, no.~18, May 2003.
\bibitem{aStephenPR69A}M.~J.~Stephen, ``Noise in the ac Josephson Effect,'' Phys.~Rev., vol.~182, no.~2, pp.~531-538, Feb 1969.
\bibitem{aStephenPR69B}M.~J.~Stephen, ``Noise in a driven Josephson oscillator,'' Phys.~Rev., vol.~186, no.~2, pp.~393-397, May 1969.



%Noise model for JJ
%RCSJ model


\bibitem{aRegalNatPhys08}C.~A.~Regal, J.~D.~Teufel, and K.~W.~Lehnert, ``Measuring nanomechanical motion with a microwave cavity interferometer,'' Nature Physics, vol.~4, no.~7, pp.~555-560, May 2008.

\end{thebibliography}

\end{document}